\documentclass[twocolumn,showpacs,floatfix,superscriptaddress,nofootinbib,amssymb,amsmath,aps,prc,eqsecnum]{revtex4}
\usepackage[dvips]{graphicx}
\usepackage{latexsym,epsfig,bm,times,psfrag,subfigure}
\usepackage{color}
\usepackage{dcolumn}                 
\usepackage[bookmarksnumbered,bookmarksopen,colorlinks,citecolor=blue,linkcolor=blue]{hyperref} %

\renewcommand\a{\alpha}
\renewcommand\b{\beta}
\renewcommand\d{\delta}
\renewcommand\k{\kappa}

\renewcommand\r{\rho}

\newcommand\e{\epsilon}
\newcommand\g{\gamma}

\newcommand\p{\pi}
\newcommand\h{\theta}

\newcommand\f{\phi}
\newcommand\w{\eta}

\renewcommand\H{\Theta}
\newcommand\D{\Delta}

\newcommand\F{\Phi}

\newcommand{\fig}[1]{Fig.~\ref{#1}}
\newcommand{\eq}[1]{Eq.~(\ref{#1})}
\newcommand{\eqs}[2]{Eqs.~(\ref{#1})-(\ref{#2})}

\newcommand\ls{\left[}
\newcommand\rs{\right]}

\newcommand{\lan}{\langle}
\newcommand{\ran}{\rangle}

\newcommand{\non}{\nonumber\\}

\newcommand{\br}{{\mathbf r}}

\newcommand{\bv}{{\bf v}}

\newcommand{\bB}{{\bf B}}
\newcommand{\bE}{{\bf E}}

\renewcommand{\part}{{\rm part}}

\begin{document}

\title{Charge-dependent azimuthal correlations from AuAu to UU collisions}
\author{John Bloczynski}
\affiliation{Physics Department and Center for Exploration of Energy and Matter,
Indiana University, 2401 N Milo B. Sampson Lane, Bloomington, IN 47408, USA.}

\author{Xu-Guang Huang}
\affiliation{Physics Department and Center for Field Theory and Particle Physics, Fudan University, Shanghai 200433, China.}
\affiliation{Physics Department and Center for Exploration of Energy and Matter,
Indiana University, 2401 N Milo B. Sampson Lane, Bloomington, IN 47408, USA.}

\author{Xilin Zhang}
\affiliation{Institute of Nuclear and Particle Physics and Department of
Physics and Astronomy, Ohio University, Athens, OH 45701, USA.}

\author{Jinfeng Liao}
\affiliation{Physics Department and Center for Exploration of Energy and Matter,
Indiana University, 2401 N Milo B. Sampson Lane, Bloomington, IN 47408, USA.}
\affiliation{RIKEN BNL Research Center, Bldg. 510A, Brookhaven National Laboratory, Upton, NY 11973, USA.}

\date{\today}

\begin{abstract}
We study the charge-dependent azimuthal correlations in relativistic heavy ion collisions, as motivated by the search for the Chiral Magnetic Effect (CME) and the investigation of related background contributions. In particular we aim to understand how these correlations induced by various proposed effects evolve from collisions with AuAu system to that with UU system. To do that, we quantify the generation of magnetic field in UU collisions at RHIC energy and its azimuthal correlation to the matter geometry using event-by-event simulations. Taking the experimental data for charge-dependent azimuthal correlations from AuAu collisions and extrapolating to UU with reasonable assumptions, we examine the resulting correlations to be expected in UU collisions and compare them  with recent STAR measurements. Based on such analysis we discuss the viability for explaining the data with a  combination of the CME-like and flow-induced contributions.
\end{abstract}


\maketitle

\section {Introduction}\label{intro}
The relativistic heavy ion collisions provide the unique way to create a new state of hot deconfined QCD matter known as the quark-gluon plasma (QGP). To measure and understand the properties of this QGP and its transition to the ordinary confined hadronic matter, being part of the early cosmos evolution history, is of fundamental interest. Such experiments are now carried out with a variety of collisional beam energies at the Relativistic Heavy Ion Collider (RHIC) as well as the Large Hadron Collider (LHC) ~\cite{Jacobs:2007dw,RHIC_WP,Muller:2012zq}. Among many other observables, two-(and multi-)particle azimuthal correlations (i.e. how particles emitted along one direction on the transverse plane get related with particles emitted along another direction) play key roles. For example, practically all collective flow measurements are done through such correlations~\cite{Voloshin:2008dg}. Other examples include, e.g. hard-soft di-hadron correlations which, via  contrast between dA and AA, were instrumental in establishing the jet quenching phenomenon~\cite{RHIC_WP}.

In the past several years, there have been significant interests in measuring and understanding the charge-dependent azimuthal correlations, that is, to study the azimuthal correlations separately for same-sign pairs  (with ++ or -- charges) and for opposite-sign pairs (with one + and the other -) and to see their differences. A major motivation came from proposals in search of possible anomalous effects, such as the  Chiral Magnetic Effect (CME)~\cite{Kharzeev:2004ey,Kharzeev:2007tn,arXiv:0711.0950,Fukushima:2008xe,Kharzeev:2009fn}, Chiral Separation Effect (CSE)~\cite{Son:2004tq,Metlitski:2005pr,Gorbar:2011ya}, Chiral Magnetic Wave (CMW)~\cite{Kharzeev:2010gd,Burnier:2011bf,Burnier:2012ae,Lin:2013sga,Hongo:2013cqa,Taghavi:2013ena,Yee:2013cya}, Chiral Electric Separation Effect (CESE)~\cite{Huang:2013iia,Jiang:2014ura}, etc. Such effects arise from nontrivial interplay between chiral fermions and QCD topological objects that are abundant in hot QGP~\cite{Liao:2006ry}, and can be manifested as generation of vector and/or axial currents in response to external strong (electro)magnetic (EM) fields. The heavy ions (such as Au, U, or Pb nuclei) with large positive electric charges and moving at nearly speed of light, naturally provide the extremely strong EM fields (on the order of hadronic scales $eE,eB\sim m_\pi^2$) at the early moments in relativistic heavy ion collisions~\cite{Rafelski:1975rf,arXiv:0711.0950,arXiv:0907.1396,arXiv:1003.2436,arXiv:1103.4239,arXiv:1107.3192,arXiv:1111.1949,arXiv:1201.5108,Bloczynski:2012en,McLerran:2013hla,Tuchin:2013ie}.  A lot of works have been done in the past few years to hunt for these effects, especially the CME, by measuring the charge-dependent azimuthal correlations and analyzing their implications~\cite{Voloshin:2004vk,Abelev:2009ac,Abelev:2009ad,Ajitanand:2010rc,Selyuzhenkov:2011xq,Abelev:2012pa}. The precise interpretations of data and the contributions of various related background effects are still under intensive investigations ~\cite{Schlichting:2010qia,Pratt:2010zn,Bzdak:2009fc,Bzdak:2010fd,Liao:2010nv,Wang:2009kd}. For recent reviews see e.g.~\cite{Bzdak:2012ia,Kharzeev:2013ffa,Liao:2014ava}.

In this paper we focus on the CME motivated measurements, for which the following charge-charge azimuthal correlations were proposed~\cite{Voloshin:2004vk} and measured both at RHIC~\cite{Abelev:2009ac,Abelev:2009ad,Ajitanand:2010rc} and at LHC~\cite{Selyuzhenkov:2011xq,Abelev:2012pa},
\begin{eqnarray}
\g_{\a\b}=\lan \cos(\f_i+\f_j-2\psi_{\rm RP})\ran_{\a\b}
\end{eqnarray}
with $\a,\b=\pm$ and $\f_i$ and $\f_j$ being the azimuthal angles of two final state charged hadrons.
The average $\lan \cdots\ran_{\a,\b}$ means first averaging over all pairs $(\f_i,\f_j)$ that consist of one $\a$ charged and one $\b$ charged hadrons and then averaging over all the events. Such azimuthal correlations are reaction-plane dependent and essentially measure the in-plane/out-of-plane difference of same-sign(SS) or opposite-sign(OS) pair correlations. The CME predicts a vector current and a resulting final hadron charge separation along the external magnetic field direction which is approximately out-of-plane. Therefore if CME is the only source contributing to these correlations, one expects the same-charge correlation ($\g_{\rm SS}$) to be negative and the opposite-charge correlation ($\g_{\rm OS}$) to be positive, with the two having the same magnitude. The STAR data for $\g_{\rm SS}$ and $\g_{\rm OS}$ indeed shows the signs and centrality trends in accord with CME expectation. However in the complex environment of heavy ion collisions, there are other
sources that could also contribute to these measured correlations. The existence of those ``background effects'' is best shown by another type of (reaction-plane independent) azimuthal correlations which were also measured~\cite{Abelev:2009ac,Abelev:2009ad,Selyuzhenkov:2011xq,Abelev:2012pa}:
\begin{eqnarray}
\d_{\a\b}=\lan \cos(\f_i-\f_j)\ran_{\a\b}.
\end{eqnarray}
As the analysis in ~\cite{Bzdak:2009fc} has shown,
while the CME-induced same-charge pair correlation (in co-moving pattern) would contribute positively to the above observable, the data show strongly negative $\d_{SS}$ which means the dominant same-sign pair correlation would be in back-to-back pattern. These measurements together strongly suggest that there are  non-CME ``background" contributions,  some of which have been identified recently, including e.g. the transverse momentum conservation (TMC) and the local charge conservation (LCC). In peripheral collision, the transverse momentum
conservation (TMC)~\cite{Pratt:2010zn,Bzdak:2010fd} tends to give both same-charge and opposite-charge correlations a negative shift which is proportional to the elliptic fow
$v_2$ and inversely proportional to multiplicity. The local charge conservation (LCC)~\cite{Schlichting:2010qia} is another source: if the charges are forced to be neutralized over  small domains in the fireball at freeze-out, then the collective flow will translate such spatial correlation into final (co-moving) moment correlation for opposite-charge pairs. Furthermore the elliptic flow will induce an in-plane/out-of-plane difference, thus giving a positive contribution to $\g_{\rm OS}$ (while no contribution to $\g_{\rm SS}$). This LCC-induced contribution originates from considering the fluid cells as canonical rather than grand-canonical ensemble of charges and such effect is also  approximately inversely proportional to multiplicity.

In addition to the known (and potentially unknown) background effects, the CME-induced signals may also suffer from  the strong initial state fluctuations of the matter geometry as well as the $\bB$-field
orientation. The event-by-event fluctuations generally make the $\bB$ direction unaligned with the event-wise second-harmonic participant plane and hence tend to suppress the CME contributions to the $\g_{\a \b}$, as thoroughly studied in ~\cite{Bloczynski:2012en}. For
RHIC AuAu collisions it was found ~\cite{Bloczynski:2012en} that such suppression is  significant for very central and very  peripheral collisions  and a strong correlation between the $\bB$ direction and matter geometry   occurs only for centrality bins around $20-50\%$. This suggests that the $\bB$-field induced effect is extremely hard to be detected in very central and very peripheral events but most likely to be detectable in the centrality class around $20-50\%$.

Clearly the present situation concerning the interpretations of the measured charge-dependent azimuthal correlations calls for a careful separation of the flow-driven background contributions. Note that both TMC and LCC effects' contributions to $\g_{\a \b}$ grows with the elliptic flow (i.e. increasing from central to peripheral collisions), however the $\bB$ field strength also shows a similar centrality trend, thus making the separation rather difficult. There have been a few different proposals in attempt to achieve this~\cite{Liao:2010nv,Bzdak:2012ia,Bzdak:2011np,Voloshin:2010ut}, one of which is to utilize the UU collisions~\cite{Voloshin:2010ut}. The Uranium nucleus $^{238}$U, unlike the Au or Pb, has a highly deformed prolate shape with a large quadrupole. The initial idea is that for the very central UU collisions there will be negligible magnetic field but still sizable geometric anisotropy (e.g. due to the so-called ``body-body'' collision configuration) that leads to elliptic flow, so that any $\bB$-related effect will be absent while any $v_2$-driven effect will still be present. Very recently, STAR collaboration reported their preliminary
results of the charge-dependent azimuthal correlations in UU collisions at $\sqrt{s}=193$ GeV~\cite{Wang:2012qs}.
The UU data show certain interesting features different from AuAu, and in particular for the most central events ($0-1\%$) they observe a sizable $v_2$ but vanishing $\g_{\rm OS}-\g_{\rm SS}$.

The meanings of these UU data, in connection with the AuAu data, require a careful examination of how various known sources of such correlations evolve from the AuAu system to the UU system. This is the main purpose of the present study. Our strategy will be to quantify the evolutions in the matter geometry (that drives flow) as well as in the EM fields from AuAu to UU, and then based on plausible assumptions to extrapolation different effects ($v_2$-related and $\bB$-related) accordingly from AuAu to UU, and see if the correlations for both systems could be consistently understood as a combination of different effects.  The rest of this paper is organized as follows. In Sec.~\ref{emfield} we will report the event-by-event calculation of the strength of the EM fields in UU collisions and the azimuthal correlations between $\bB$-field orientation and the event-wise participant plane. In Sec.~\ref{exper} we will then use these results to extrapolate the decomposed ($v_2$-related and $\bB$-related) components  of the correlations $\g_{\a\b}$ from AuAu data to UU collisions and compare them with UU data. Finally we summarize in Sec.~\ref{discu}. The natural unit $\hbar=c=k_B=1$ will be used throughout this paper.

\section {Electromagnetic fields in UU collisions with initial state fluctuations}\label{emfield}

The possibility to study heavy ion collisions using UU system has been discussed for long, see e.g. \cite{Shuryak:1999by,Li:1999bea,Heinz:2004ir,Masui:2009qk,Filip:2009zz,Haque:2011aa}. Event-by-event simulations were also previously done to study the expectations for multiplicities as well as the geometric anisotropies in the initial conditions for UU collisions. The magnitude of strong electromagnetic fields, crucial for those field-induced effects in UU collisions, was first estimated in~\cite{Voloshin:2010ut} with the approximation of counting all spectator contributions. To develop future sophisticated modelings that may be compared with data, however, it would be important have a full analysis of the field strength and even more importantly the azimuthal correlations between the fields and the matter geometry, which is still lacking. In this Section, we report our event-by-event determination of such EM fields and particularly their azimuthal orientations with respect to the (concurrently fluctuating) initial matter geometry.

\subsection {Setup}\label{setup}
We first discuss our setup for the   event-by-event analysis of the electromagnetic fields generated in
UU collision at $\sqrt{s}=193$ GeV, following similar simulations done for AuAu collisions in ~\cite{Bloczynski:2012en}. Let us focus on the fields at the initial time, $t=0$, that is,
the time when the centers of the two colliding uranium nuclei both lie on the transverse plane. The time dependence of these fields will have no difference from that in AuAu case as previously studied (see e.g. \cite{arXiv:1201.5108,McLerran:2013hla}). Throughout this Section we will show results for the field point at the center of fireball $x=y=z=0$. We've also computed the fields at other transverse points for UU collisions and found the dependence on the transverse position is  similar to that in the AuAu case~\cite{Bloczynski:2012en}.
We utilize the following form of the
Li\'enard-Wiechert potentials for constantly moving charges:
\begin{eqnarray}
\label{LWE}
e\bE(t,\br)&=&\frac{e^2}{4\p}\sum_n Z_n({\bf R}_n)\frac{1-v_n^2}{[R_n^2-({\bf R}_n\times\bv_n)^2]^{3/2}}{\bf R}_n, \non \\
\label{LWB}
e\bB(t,\br)&=&\frac{e^2}{4\p}\sum_n Z_n({\bf R}_n)\frac{1-v_n^2}{[R_n^2-({\bf R}_n\times\bv_n)^2]^{3/2}}{\bf v}_n\times{\bf R}_n,\non
\end{eqnarray}
where ${\bf R}_n=\br-\br_n(t)$ is the relative position of the field point $\br$ to the $n$th proton at time $t$, $\br_n(t)$, and $\bv_n$ is the velocity of the $n$th proton.
The summations run over all protons in the projectile and target nuclei.
Equations (\ref{LWE}) and (\ref{LWB}) contain
singularities at $R_n=0$ if we treat protons as point charges. In practical
calculation, to avoid such singularities we treat protons as uniformly charged spheres with radius $R_p$. The charge number factor $Z_n(\bf{R}_n)$ in Eqs.~(\ref{LWE}) and (\ref{LWB}) is introduced to encode this aspect: when the field point locates outside the $n$th proton (in the rest frame of the proton) $Z_n=1$, otherwise $Z_n<1$ depends on ${\bf R}_n$. The in-medium charge radius $R_p$ of proton is unknown, thus we
use the vacuum value $R_p=0.84184(67)$ fm ~\cite{Pohl:2010zza} in our numerical simulation.  Varying $R_p$ from $0.6$ fm to $0.9$ fm brings about $15\%$ variation in the computed field strength.

The nucleons in one nucleus move at constant velocity along the beam direction (we choose it as $z$-direction) while the nucleons
in the other nucleus move at the same speed but opposite direction. The energy for each nucleon is set to be $\sqrt{s}/2$ in the
center-of-mass frame, therefore the value
of the velocity of each nucleon is given by $v_n^2=1-(2m_N/\sqrt{s})^2$, where $m_N$ is the mass of the nucleon. We set
the $x$-axis along the impact parameter vector so that the reaction plane is the $x$-$z$ plane. Finally,
the positions of nucleons in the rest frame of a uranium nucleus are sampled according to a deformed Woods-Saxon distribution (in spherical coordinates)~\cite{Masui:2009qk,Filip:2009zz,Haque:2011aa},
\begin{eqnarray}
\r(r,\h)&=&\frac{\r_0}{1+\exp{[(r-R'(\h))/a]}},\\
R'(\h)&=&R[1+\b_2 Y_2^0(\h)+\b_4 Y_4^0(\h)],
\end{eqnarray}
where $\r_0=0.16$ fm$^{-3}$ is the normal nuclear density, $R$ and $a$ denote the ``radius" of
the nucleus and the surface diffuseness parameter. The parameters $\b_2$ and $\b_4$ specify
the shape deformation of the nucleus from sphere ($\b_2=\b_4=0$).
$Y^m_l(\h)$ denotes spherical harmonics and $\h$ is the polar
angle with respect to the symmetry axis of the nucleus. The values of the parameters for $^{238}U$
are $R = 6.81$ fm, $a = 0.55$ fm, $\b_2 = 0.28$, and $\b_4 = 0.093$.
The positions of nucleons are sampled
by $4\p r^2\sin\h\r(r)d\h d\f$ with appropriate normalization. Different though from the AuAu collisions, in the UU case the polar and azimuthal directions of the rotation-symmetric axis for each of the colliding uranium nuclei are randomly (and independently) orientated on the event-by-event basis with probability density $\sin\H$ and uniform
distribution for $\H$ and $\F$, respectively. Here, $\H$ and $\F$ are the solid angles of the rotation-symmetric axis of each  nucleus. The $\sin\H$ weight needs to be implemented to simulate unpolarized nucleus-nucleus
collisions. For each given impact parameter we sample 100 different orientational configurations.

\subsection {Electromagnetic fields}\label{field}
\begin{figure}
\begin{center}
\includegraphics[width=6.5cm]{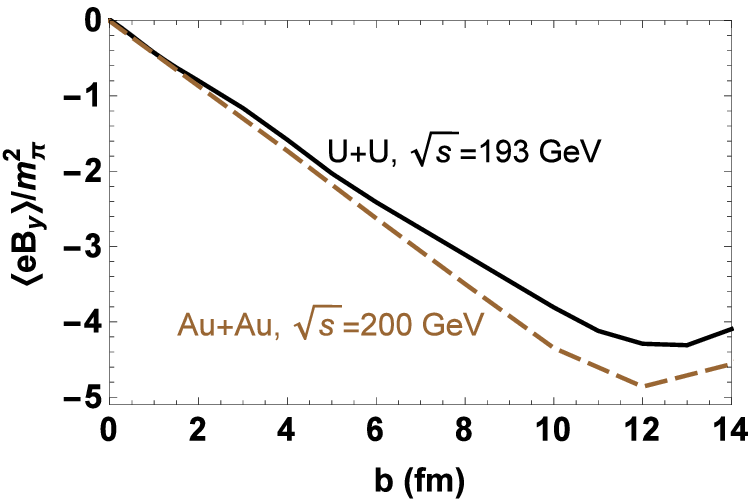}
\caption{The event-averaged $eB_y$(in unit of $m^2_\pi$) at $t=0$
as a function of the impact parameter $b$.}
\label{fig_by}
\end{center}
\end{figure}

In \fig{fig_by}, we show the result for the event-averaged $y$-component of the magnetic field, $\lan eB_y\ran$, generated in UU collision at $\sqrt{s}=193$ GeV. (The negative values have no particular meaning, only related to choice of which nucleus is running which way from beam pipe view.) Despite the fact that the shape of uranium nucleus is rather different from
sphere, after event average, its effects in generating electromagnetic fields are nearly equivalent to a spherical nucleus. Indeed, it is seen from \fig {fig_by} that the magnitude and the impact parameter dependence of $\lan eB_y(b)\ran$ in UU collisions are just slightly smaller than that in AuAu collision at $\sqrt{s}=200$ GeV (as reported in e.g.  \cite{arXiv:1111.1949,arXiv:1201.5108}). Other components, $\lan eB_{x,z}\ran$, of the magnetic field and the electric field $\lan e\bE\ran$
are essentially zero after event average.

\begin{figure}
\begin{center}
\includegraphics[width=6.5cm]{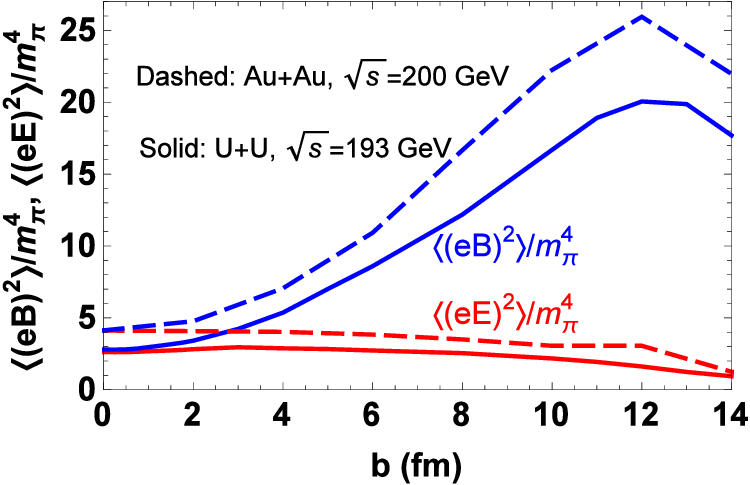}
\caption{(Color online) The event-averaged $(eB)^2$ and $(eE)^2$ (in unit of $m^4_\pi$) at $t=0$
as functions of the impact parameter $b$.}
\label{fig_field2}
\end{center}
\end{figure}

As already shown in the AuAu and PbPb cases~\cite{Bloczynski:2012en}, the event-by-event position fluctuation of the protons in
the colliding nuclei can bring interesting features, for example,
a large field strength square, even for central collision.
The similar situation happens for UU collision, see
\fig{fig_field2} for the event-averaged magnetic field square and electric field square, $\lan(e\bB)^2\ran$ and $\lan(e\bE)^2\ran$. We see that on the event-by-event basis, the magnitudes of $\lan(e\bB)^2\ran$ and $\lan(e\bE)^2\ran$ are smaller than that in AuAu collisions reported in~\cite{Bloczynski:2012en}: the nucleus shape matters in this sense.

\begin{figure}
\begin{center}
\includegraphics[width=6.5cm]{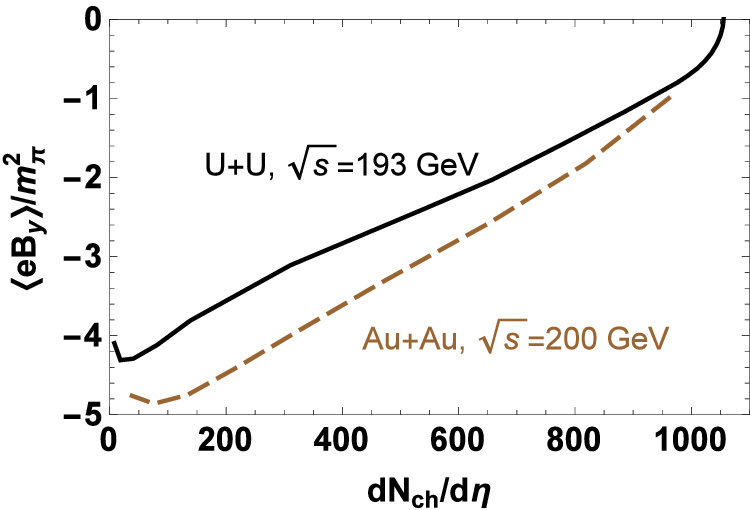}
\caption{The event-averaged $eB_y$(in unit of $m^2_\pi$) at $t=0$
as a function of the multiplicity.}
\label{fig_by_mul}
\end{center}
\end{figure}

\begin{figure}
\begin{center}
\includegraphics[width=6.5cm]{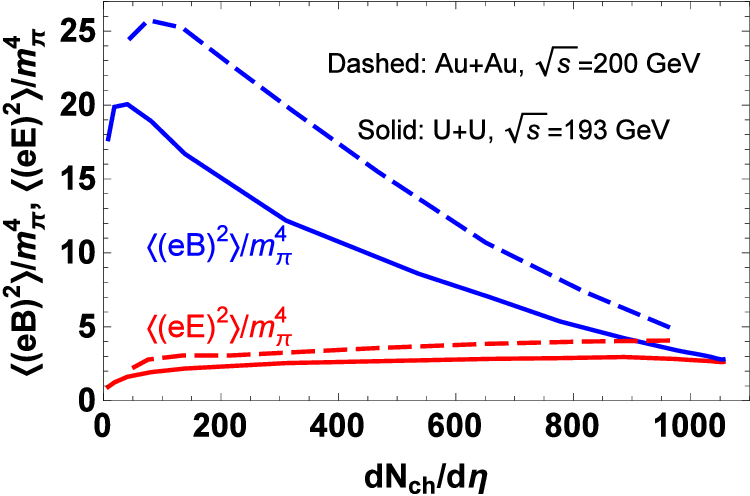}
\caption{(Color online) The event-averaged $(eB)^2$ and $(eE)^2$ (in unit of $m^4_\pi$) at $t=0$
as functions of the multiplicity.}
\label{fig_field2_mul}
\end{center}
\end{figure}

For the convenience of comparison with experimental data later, we also plot $\lan eB_y\ran$ and $\lan(e\bB)^2\ran$ and $\lan(e\bE)^2\ran$ as functions of the charged particle multiplicity per
unit pseudorapidity (we will simply call it multiplicity)
near mid-rapidity region. We adopt the following two component model~\cite{Kharzeev:2000ph}
\begin{eqnarray}
\frac{d N_{ch}}{d\w}=n_{pp}\ls(1-x)\frac{N_{\rm part}}{2}+x N_{\rm coll}\rs
\end{eqnarray}
to describe the multiplicity as functions of participant number $N_{\rm part}$ and binary collision number $N_{\rm coll}$, where $n_{pp}\approx 2.42$ is the energy-dependent charge multiplicity of
proton-proton collision at $\sqrt{s}=193$ GeV and $x=0.13$ is the fraction of
charged multiplicity generated in binary nucleon-nucleon collisions~\cite{wanggang}. So the event-wise multiplicity can be determined from $N_{\rm part}$ and $N_{\rm coll}$ in each event. Note that multiplicity fluctuations may be particularly important for the analysis of most central collisions as studied in e.g. ~\cite{Voloshin:2010ut}.
The multiplicity dependence of $\lan eB_y\ran$ and $\lan(e\bB)^2\ran$ and $\lan(e\bE)^2\ran$ are shown in \fig{fig_by_mul} and \fig{fig_field2_mul}.

\subsection {Azimuthal correlation between the $\bB$-field orientation and the participant planes}\label{azimu}
We now turn to study the azimuthal correlation between the magnetic field and the participant plane, i.e., we will calculate $\lan\cos[2(\psi_{\bf B}-\psi_2)]\ran$ and
$\lan(e\bB)^2\cos[2(\psi_{\bf B}-\psi_2)]\ran/\lan(e\bB)^2\ran$ where $\psi_\bB$ is the azimuthal
angle of the magnetic field and $\psi_2$ is the azimuthal angle of the second harmonic participant plane. The motivation and importance of such a study has been explained in
~\cite{Bloczynski:2012en}: basically the $\bB$-field-induced effect will bear such a suppression factor, so the quantitative contribution of such an effect is controlled not by $\lan(e\bB)^2\ran$ but by $\lan(e\bB)^2\cos[2(\psi_{\bf B}-\psi_2)]\ran$ which may be called the ``projected field strength''. Let us mention in passing that the matter geometry itself is fluctuating, and with our event-by-event simulations we determine the harmonic participant planes for each event  from the Monte Carlo Glauber simulations of the initial condition and analyze the angular correlations between
the $\bB$ and the participant plane orientations from the same event. The $n$th harmonic participant plane angle $\psi_n$ and eccentricity $\e_n$ are calculated from participant density $\r(\br)$ as in the literature:
$\e_1e^{i\psi_1} =-(\int d^2\br_\perp\r(\br_\perp) r_\perp^3 e^{i\f})/({\int d^2\br_\perp\r(\br_\perp) r_\perp^3}),$ and
$\e_n e^{in\psi_n}=-({\int d^2\br_\perp\r(\br_\perp) r_\perp^n e^{in\f})/(\int d^2\br_\perp\r(\br_\perp) r_\perp^n})$ for $n>1$.

\begin{figure}
\begin{center}
\includegraphics[width=6.5cm]{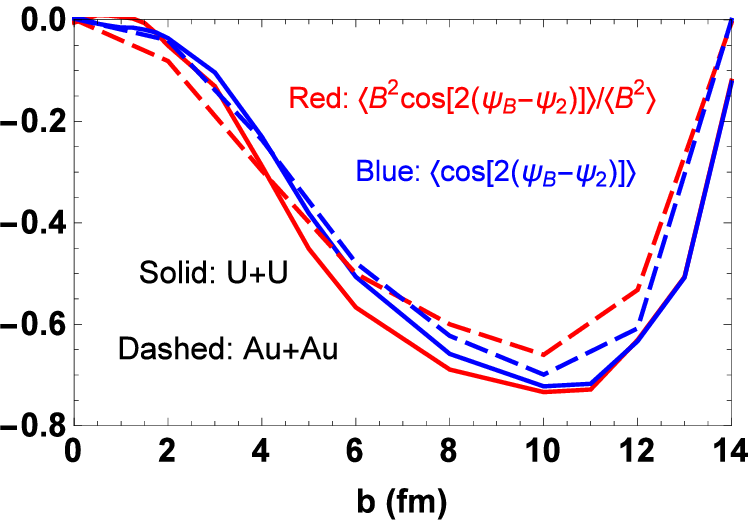}
\caption{(Color online) The correlations $\lan\cos[2(\Psi_{\bf B}-\Psi_2)]\ran$ and $\lan(e\bB)^2\cos[2(\Psi_{\bf B}-\Psi_2)]\ran/\lan(e\bB)^2\ran$ as functions of impact parameter $b$.}
\label{correlb}
\end{center}
\end{figure}
\begin{figure}
\begin{center}
\includegraphics[width=6.5cm]{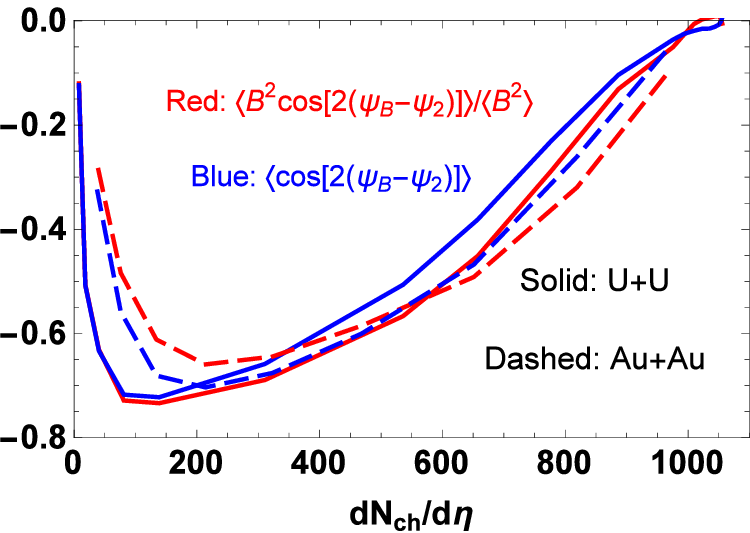}
\caption{(Color online) The correlations $\lan\cos[2(\Psi_{\bf B}-\Psi_2)]\ran$ and $\lan(e\bB)^2\cos[2(\Psi_{\bf B}-\Psi_2)]\ran/\lan(e\bB)^2\ran$ as functions of multiplicity.}
\label{correlmul}
\end{center}
\end{figure}
In \fig{correlb} and \fig{correlmul} we show these azimuthal correlations in UU collision as functions
of impact parameter and the multiplicity, respectively. For comparison, also shown are the same quantities for AuAu collisions. Again, we see that the correlations (as functions of $b$) behave similarly as that in AuAu collision: at small and large impact parameters the correlations are
strongly suppressed while the strongest correlations ($\sim 0.7$) occur around $b\approx 10$ fm or $dN_{ch}/d\w$ around $200$.

\begin{figure}
\begin{center}
\includegraphics[width=6.5cm]{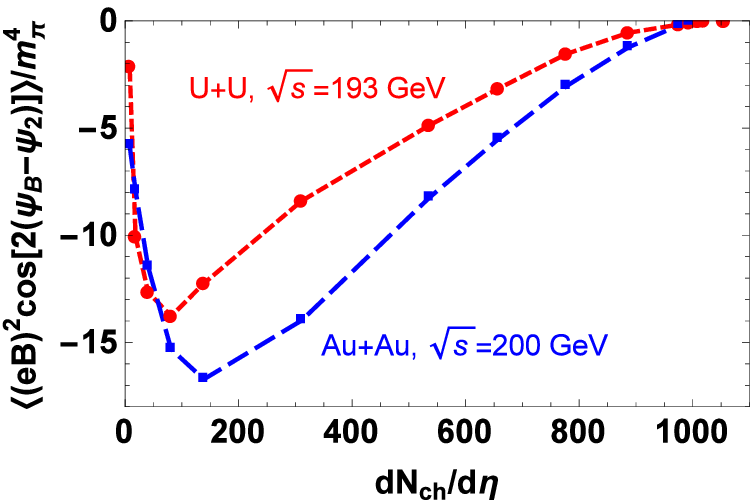}
\caption{(Color online) The ``projected field strength'' $\lan(e\bB)^2\cos[2(\Psi_{\bf B}-\Psi_2)]\ran$ as functions of the multiplicity for AuAu (blue square) and UU (red circle) collisions.}
\label{fig_b2cos}
\end{center}
\end{figure}

It is of great interest to compare the important quantity $\lan(e\bB)^2\cos[2(\psi_{\bf B}-\psi_2)]\ran$ that controls the strength of $\bB$-induced effect for both UU and AuAu collisions: see \fig{fig_b2cos}. We see that
this strength in UU collision is generally weaker than that in AuAu collision for most of the centrality except for the very central and very peripheral cases where both tend to vanish. This is mainly because the U nucleus has a bigger size than the Au nucleus and thus the spectators are further away from the field point thus  generating a smaller magnetic field magnitude in UU than in AuAu (see \fig{fig_field2}). We note a similar smaller magnitude of $\bB$ in UU than in AuAu is also shown in ~\cite{Voloshin:2010ut}.

\begin{figure}
\begin{center}
\includegraphics[width=6.5cm]{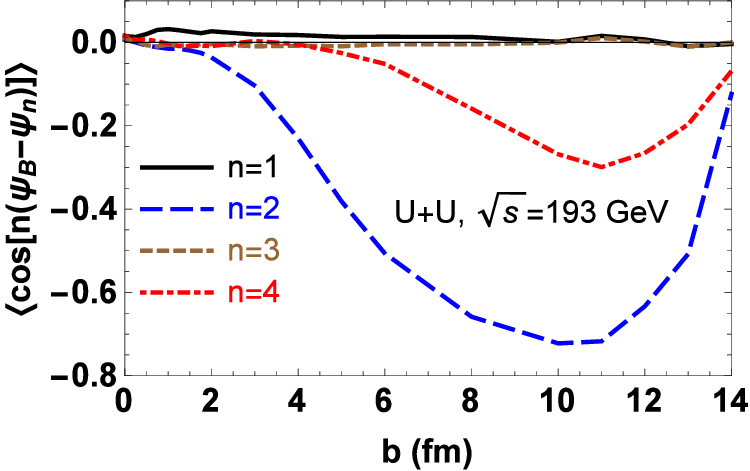}
\includegraphics[width=6.5cm]{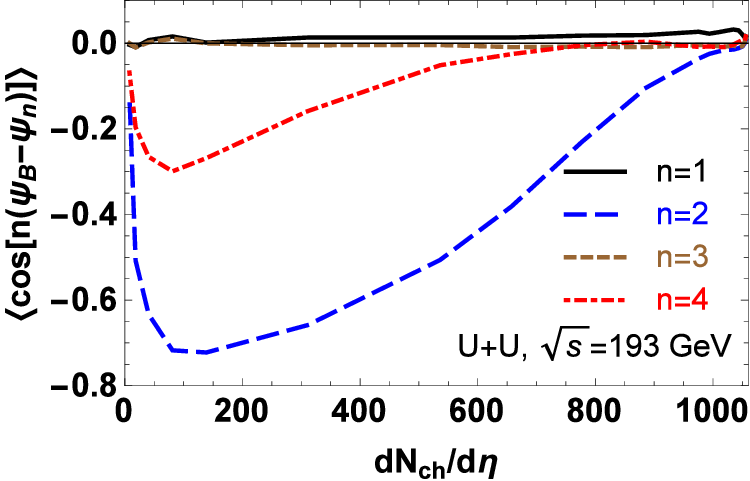}
\caption{(Color online) The correlations $\lan \cos[n(\Psi_{\bf B}-\Psi_n)]\ran$ as functions of the impact parameter and the multiplicity, respectively.}
\label{fig_cosn}
\end{center}
\end{figure}

We have also studied the azimuthal correlations between $\bB$-field and the
participant planes defined by other harmonics. In \fig{fig_cosn}
we show the $\lan\cos[n(\psi_{\bf B}-\psi_n)]\ran$ as functions of impact parameter and multiplicity, respectively. Again, we see very similar patterns as in AuAu collisions ~\cite{Bloczynski:2012en}, including the impact parameter
dependence of $n=2,4$ cases and the nearly vanishing $n=1,3$ cases.

\section {Implications on the charge-dependent azimuthal correlations}\label{exper}

In this Section we carry out an extrapolation study from AuAu to UU systems. With the event-by-event simulations done for UU above and for AuAu previously in \cite{Bloczynski:2012en}, we know how the matter geometry (i.e. eccentricity $\epsilon_2$ driving elliptic flow) changes from AuAu to UU, and we know how the $\bB$ field and its orientation (which drives effect like CME) changes from AuAu to UU. So with reasonable assumptions one may develop extrapolations on how the contributions to charge-dependent correlations from the two types of sources , the $v_2$-related and the $\bB$-related, would evolve from AuAu to UU systems. In what follows, we will first try to decompose the two types of sources in the AuAu data, and then extrapolate them respectively to UU, and see if a reasonable description of UU data would be achieved by such a combination of different effects.

\subsection {Decomposition of the charge-dependent azimuthal correlations}\label{decom}
Let us focus on the CME-motivated azimuthal correlation (which by itself is parity-even, able to measure the fluctuations of a parity-odd charge dipole)
\begin{eqnarray}
\g_{\a\b}=\lan \cos(\f_i+\f_j-2\psi_{RP})\ran_{\a\b}.
\end{eqnarray}
It is also important to simultaneously examine the other charge-dependent azimuthal correlation
\begin{eqnarray}
\d_{\a\b}=\lan \cos(\f_i-\f_j)\ran_{\a\b}.
\end{eqnarray}
To simplify the notation, we will also use $\g_{\rm SS}$ (SS for same sign) and $\g_{\rm OS}$ (OS for opposite sign) to denote $\g_{++/--}$ and $\g_{+-/-+}$ (similarly for $H_{\rm SS,OS}, F_{\rm SS,OS}, \d_{\rm SS,OS}$ below), respectively.

The CME-induced charge separation along the out-of-plane direction (see e.g. discussions in ~\cite{Bzdak:2009fc,Bzdak:2012ia}), would make the following contributions to these correlations:
\begin{eqnarray}
\g^{\rm CME}_{\a\b}\equiv -H_{\a\b}  \,\, , \,\,  \d^{\rm CME}_{\a\b}=H_{\a\b}   \, .
\end{eqnarray}
with $H_{SS} \simeq -H_{OS}>0$.

\begin{figure*}
\begin{center}
\includegraphics[width=6.5cm]{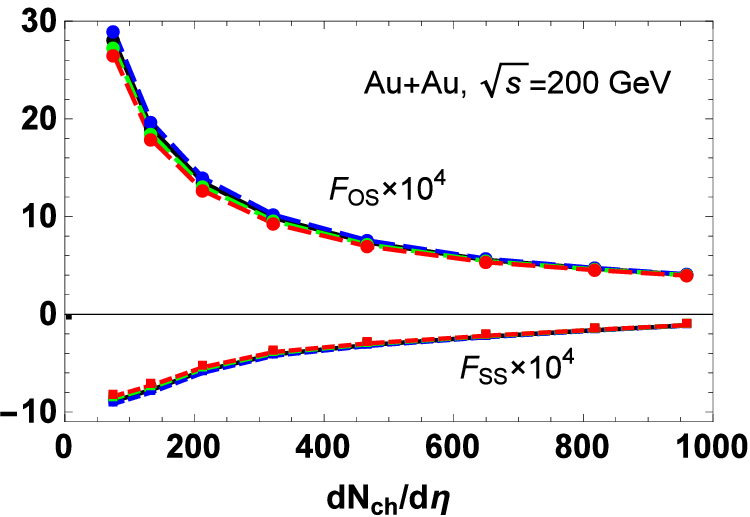}
\includegraphics[width=6.5cm]{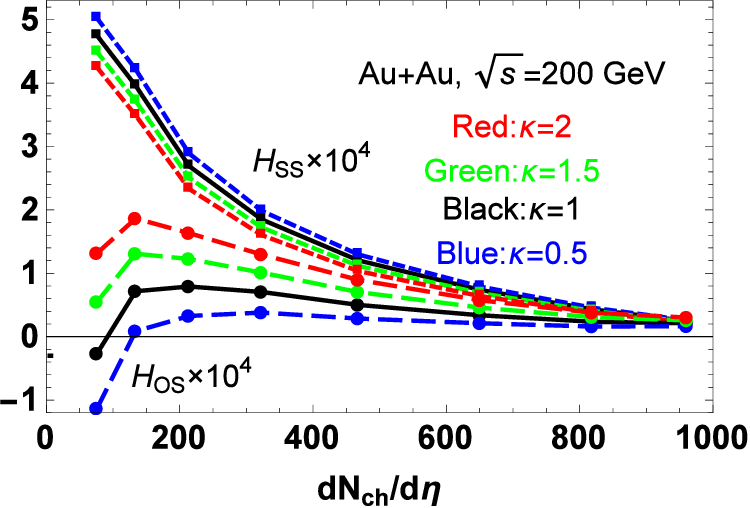}
\caption{(Color online) Functions $F_{\rm SS,OS}$ and $H_{\rm SS, OS}$ versus multiplicity, extracted from the decomposition analysis in Eqs.(\ref{gab})(\ref{dab}) of the STAR AuAu data. Different colors correspond to different choices of $\k$.}
\label{FH}
\end{center}
\end{figure*}

\begin{figure*}
\begin{center}
\includegraphics[width=6.5cm]{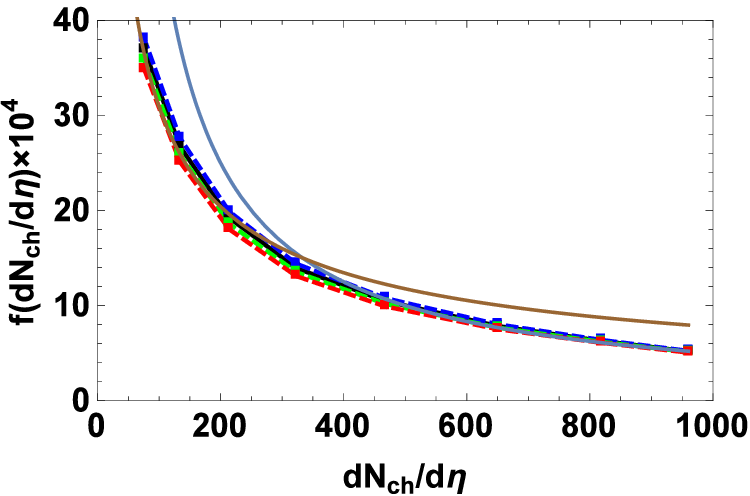}
\includegraphics[width=6.5cm]{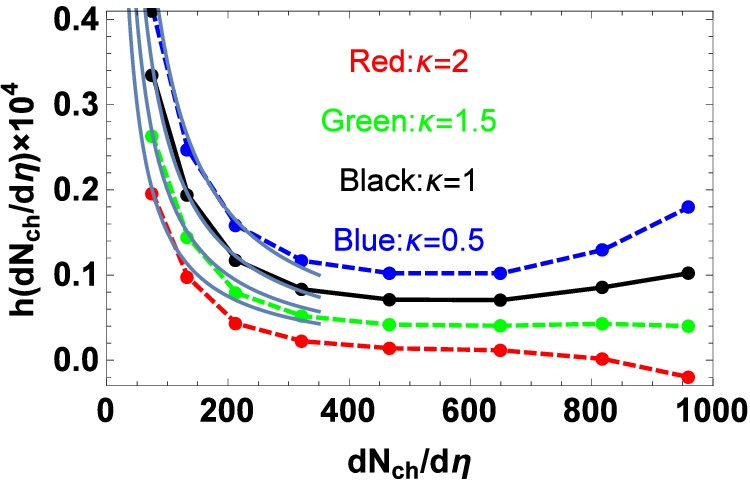}
\caption{(Color online) The extracted functions $f[{dN_{ch}}/{d\w}]$ and $h[{dN_{ch}}/{d\w}]$ (introduced in Eqs.(\ref{dgamma}) and (\ref{ddelta})) versus multiplicity, with the thin curves being fitting results (see text for details). Different colors correspond to different choices of $\k$.}
\label{fh}
\end{center}
\end{figure*}

As already discussed in the Introduction, there  are non-CME ``background" contributions to
both $\g$ and $\d$, including the  transverse momentum conservation (TMC) and the local charge conservation (LCC). The dynamical mechanisms of both effects are intrinsically reaction-plane independent, with TMC inducing a charge-independent back-to-back correlation while the LCC inducing a near-side (co-moving) correlation solely in the opposite-charge pair correlation. They however contribute to the reaction-plane dependent correlation $\g$ due to the anisotropy in the particle azimuthal distribution that is quantified by $v_2$. As briefly analyzed in ~\cite{Bzdak:2012ia}, the contributions of these effects to the $\d$ and the $\g$ are related roughly by a factor of $v_2$.

One can therefore make the following plausible decomposition of the two types, CME-like and $v_2$-related, contributions to these correlations in the following two-component way:
\begin{eqnarray}
\label{gab}
\g_{\a\b}&=&\k v_2 F_{\a\b}-H_{\a\b},\\
\label{dab}
\d_{\a\b}&=&F_{\a\b}+H_{\a\b},
\end{eqnarray}
where $F_{\a\b}$ are the $v_2$-related background contributions, $H_{\a\b}$ are the CME contributions. Note that we have introduced an ``uncertainty factor'' $\k$ in the above, following the approach of the recent STAR analysis published in ~\cite{Adamczyk:2014mzf}. While in certain ideal limit $\k$ would be one, there are practical issues such as  finite kinematic selections in real experiments that may drive $\k$ to deviate from unity~\cite{Adamczyk:2014mzf}. In order to account for such uncertainties, we will vary the factor with four different values $\k=0.5, 1, 1.5$, and $2$ for our subsequent numerical analysis and see how the results change with such choices. This shall give a reasonable idea about the impact of such uncertainties.

With such working assumptions, we
can then use the STAR AuAu collision data for $\g_{\a\b}, \d_{\a\b}$~\cite{Abelev:2009ac,Abelev:2009ad,Wang:2012qs}, and $v_2$ to extract $F_{\a\b}$
and $H_{\a\b}$ as functions of centrality or multiplicity for a given $\k$. The result of such a decomposition is shown in
\fig{FH}. In the decomposition, we have used $v_2\{\eta\; {\rm sub}\}$~\cite{Adams:2004bi}; using other measurements of $v_2$,
like $v_2\{\rm EP\}$, $v_2\{2\}$ gives minor difference in $h$ and $f$ below.
From the plots one can see that: 1) the flow-related parts, $F_{OS,SS}$ (including their centrality trends) can be reasonably understood as a combination of TMC and LCC, with TMC  making the negative contributions to both the $F_{SS}$ and $F_{OS}$ while the LCC making a strong positive contribution to the opposite sign correlations accounting for the difference between $F_{SS}$ and $F_{OS}$; 2) the remaining correlations (that are unrelated to $v_2$) in the $H$ still show charge-dependence and the $H_{SS}$ appears in line with the CME expectation; 3) the $H_{OS}$ for most centrality is positive (which would be different from the pattern induced by  CME alone), and a plausible explanation of $H_{OS,SS}$ together could be that a charge-independent effect (such as the matter dipole fluctuations pointed out in \cite{Teaney:2010vd}) contributes to both while the CME accounts for the difference between $H_{OS}$ and $H_{SS}$. 4) the $\k$-dependence of $F_{OS/SS}$ are minor, while $H_{OS/SS}$ appears to be sensitive to the choice of $\k$ values. For smaller $\k$, the difference between $H_{OS}$ and $H_{SS}$ becomes more significant. Lastly, it is worth emphasizing that a single component assumption with only flow-related contributions would {\it not} work for describing all the data and a CME-like component appears necessary for a full description of full data set. Our decomposition analysis results for AuAu collisions are consistent with that in ~\cite{Adamczyk:2014mzf}.

A useful way to eliminate any charge-independent contributions and focus on the truly charge-dependent correlations is to make a subtraction between SS and OS signals,
\begin{eqnarray}
\D\g&\equiv& \g_{\rm OS}-\g_{\rm SS},\\
\D\d&\equiv& \d_{\rm OS}-\d_{\rm SS}.
\end{eqnarray}
In $\D\g$ and $\D\d$, charge independent sources are then subtracted out. Now, from \eqs{gab}{dab}, we can write
\begin{eqnarray}
\label{dgamma}
\D\g&=&\k f v_2-h\lan(e\bB)^2\cos[2(\psi_{\bf B}-\psi_2)]\ran,\\
\label{ddelta}
\D\d&=&f+h\lan(e\bB)^2\cos[2(\psi_{\bf B}-\psi_2)]\ran,
\end{eqnarray}
In the above, $f=F_{OS}-F_{SS}$ is the flow-driven component. The $h\lan(e\bB)^2\cos[2(\psi_{\bf B}-\psi_2)]\ran = H_{OS}-H_{SS}$ represents the CME-like signal in $H_{OS}-H_{SS}$ that is driven by $\bB$-field and that explicitly depends on the ``projected field strength'' along the event-wise out-of-plane direction.

Finally let us try to extract  the two functions $f$ and $g$ introduced in Eqs.(\ref{dgamma})(\ref{ddelta}), with the assumption that they are dominantly determined by multiplicity ${dN_{ch}}/{d\w}$, i.e. $f=f[{dN_{ch}}/{d\w}]$ and $h=h[{dN_{ch}}/{d\w}]$ for a given collisional beam energy (noting that the AuAu $200\, \rm GeV$ and UU $193\, \rm GeV$ could be considered approximately the same for all practical purposes). This assumption is plausible as we have explicitly separated out the other important factors $v_2$ and $\lan(e\bB)^2\cos[2(\psi_{\bf B}-\psi_2)]\ran$ in such correlations.  One can then extract these two functions from the $F_{\a\b}$ and $H_{\a\b}$ (in Fig.\ref{FH}), given the information about the ``projected field strength'' (from ~\cite{Bloczynski:2012en}): the final results for $f$ and $h$ are shown in \fig{fh}. We see that similar with that for $F_{OS/SS}$ and $H_{OS/SS}$, $f$ is almost independent of $\k$ while $h$ is sensitive to $\k$. Theoretically both functions may be expected to have a crude linear dependence on inverse multiplicity, and we have attempted to test this by a fitting analysis, with the obtained fitting curves also shown in
\fig{fh}. It is found that: (1) $f(x)\propto 1/x$ at high multiplicity region and
$\propto 1/x^{0.6}$ at low multiplicity region; (2) $h(x)\propto 1/x$ at low multiplicity region (but with different proportional factor according to the different values of $\k$) and $h$ seems roughly a constant when multiplicity is large.

\subsection {Extrapolation from AuAu to UU collisions and comparison with STAR data}\label{uustar}

We are now ready to make an extrapolation analysis from AuAu to UU collisions. Our starting point is to assume that the two decomposition relations in \eq{dgamma} and \eq{ddelta} apply to UU collisions as well, with three important ingredients being implemented in the following way: 1) the projected field strength $\lan(e\bB)^2\cos[2(\psi_{\bf B}-\psi_2)]\ran$ to be used are the values for UU system (as shown in Fig.\ref{fig_b2cos}); 2) the $v_2$ to be used are the measured values in UU collisions at given multiplicity (as measured by STAR~\cite{Wang:2012qs}); 3) the two functions $f[{dN_{ch}}/{d\w}]$ and $h[{dN_{ch}}/{d\w}]$ are assumed to remain the same as extracted from AuAu analysis and shown in Fig.\ref{fh} (with slight linear extrapolation to somewhat larger multiplicity that is only accessible in UU collisions beyond that in AuAu collisions). With these we are then able to make a reasonable ``guess'' of the azimuthal correlations $\D\g$ and $\D\d$ in UU collisions that could be compared with measured data.

\begin{figure}
\begin{center}
\includegraphics[width=6.5cm]{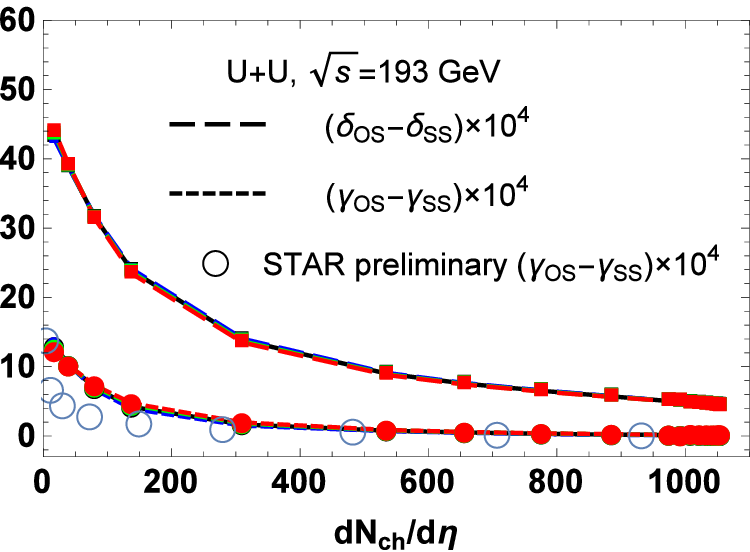}
\caption{(Color online) The extrapolation results for charge-dependent azimuthal correlations $\D\g$ and $\D\d$ in UU collisions at $\sqrt{s}=193\,  \rm GeV$ (see text for details). The open circles are STAR preliminary data~\cite{Wang:2012qs}.}
\label{fig_dgdd}
\end{center}
\end{figure}

\begin{figure}
\begin{center}
\includegraphics[width=6.5cm]{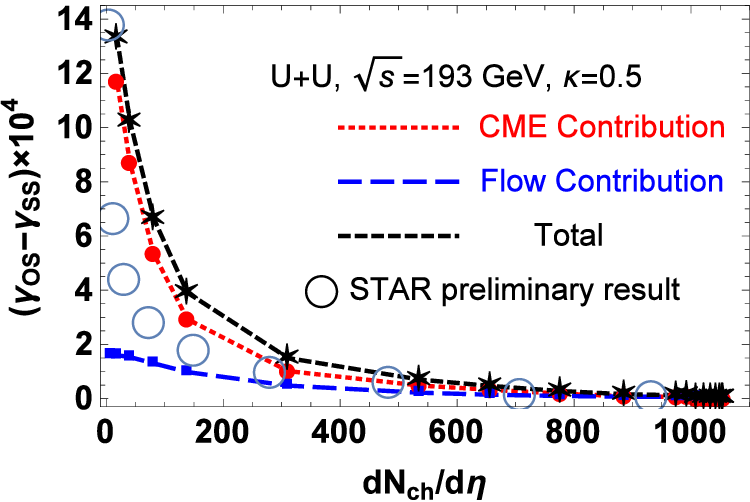}
\includegraphics[width=6.5cm]{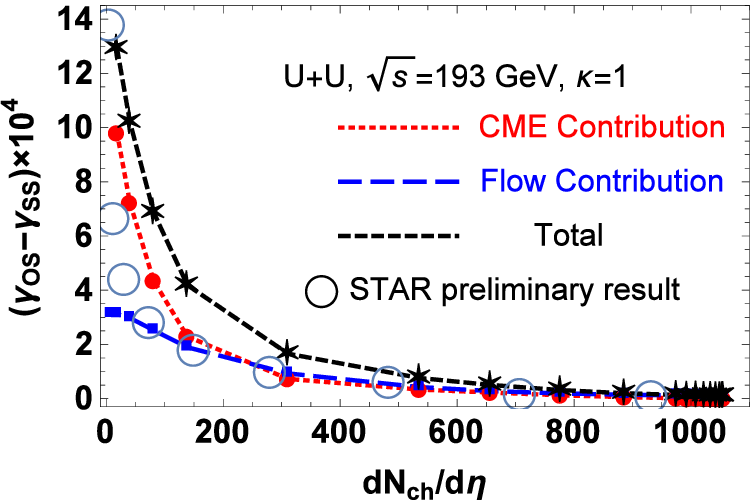}
\includegraphics[width=6.5cm]{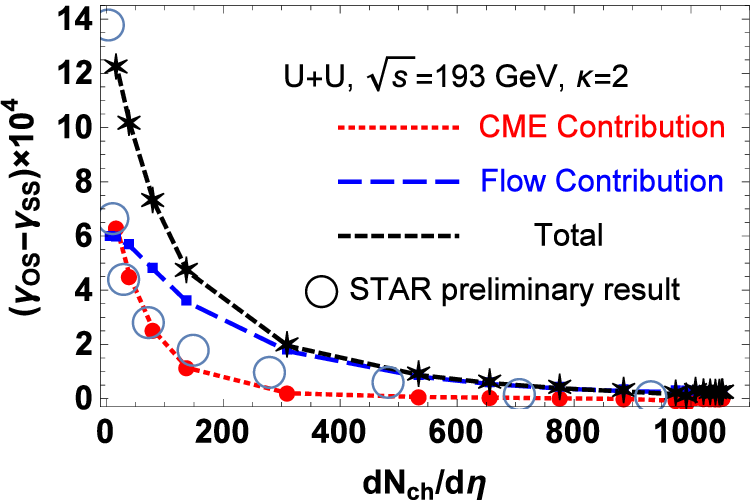}
\caption{(Color online) Contributions from different sources to the correlation $\D\g$  for UU collisions at $\sqrt{s}=193 \, \rm GeV$ (see text for details). The open circles are STAR preliminary data~\cite{Wang:2012qs}. Each panel corresponds to a specific choice of $\k$.}
\label{fig_dg2}
\end{center}
\end{figure}
In \fig{fig_dgdd} we show our results from such an extrapolation analysis for $\D\g$ and $\D\d$ in UU collisions
at $\sqrt{s}=193$ GeV. The STAR preliminary data for $\D\g$~\cite{Wang:2012qs} are also shown. As one can see, the extrapolated results are close to the data except for the low multiplicity region and are not sensitive to the choice of $\k$ in the scale depicted in this figure. $\D\d$ is dominated by $f$ and is therefore insensitive to $\k$. The $\D\g$, on the other hand, does change with the values of $\k$. To have a more detailed understanding of the contributions from the two components of sources in the correlations, we show in \fig{fig_dg2}  the  respective contributions from CME-like source and $v_2$-related source to the $\D\g$. Interestingly we find that each individual component could give a fairly reasonable account of the data but the added total is clearly exceeding the data. Furthermore, while each individual component's contribution is sensitive to the choice of $\k$, their sum is much less sensitive to the choice of $\k$ as the two change oppositely with the value of $\k$. The discrepancy is mostly in the low multiplicity region. It might indicate either the incompleteness of the two-component hypothesis or possible more complicated dependence of the two functions $f$ and $h$ than our simple extrapolation. We though emphasize that in the most interesting high multiplicity regime the extrapolation results show very reasonable agreement with data.

\begin{figure}
\begin{center}
\includegraphics[width=7cm]{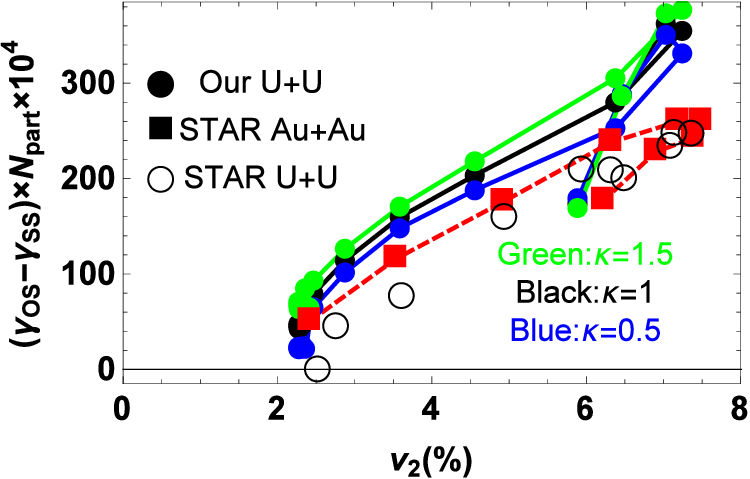}
\caption{(Color online) $\D\g$ (scaled up by factor $N_{part}$) versus $v_2$ for AuAu (circle) and UU (square) collisions. The filled symbols are from our analysis and the open symbols are STAR preliminary data~\cite{Wang:2012qs}.}
\label{fig_dgv2}
\end{center}
\end{figure}

Finally in \fig{fig_dgv2} we plot the azimuthal correlations $\D\g$ (scaled up by $N_{part}$ following convention in ~\cite{Wang:2012qs}) versus $v_2$
for both AuAu and UU collisions, in comparison with available data~\cite{Wang:2012qs}. While the general trends and order of magnitude for extrapolation results  are in crude proximity to the preliminary UU data, a quantitative agreement is lacking especially in the peripheral regime. Nevertheless, an interesting observation is that our UU curve from extrapolation is roughly parallel to the data curve and the comparison appears to favor a smaller $\k$ value. The origin of such a discrepancy is not clear at the moment. So what can one learn from the mismatch between the extrapolated results and the data for UU system? Logically there could be varied reasons: 1) it could be due to the assumption that the functions $f$ and $g$ in Fig.\ref{fh} can be simply extrapolated from AuAu to UU; 2) it could be that the way of linearly combining the two types of identified sources may not be fully accurate; 3) it could also be that the two-component scenario may not be entirely true and there could be other unknown type of sources contributing to the correlations. It however should be emphasized that there is no indication of non-existence for any of the two types of already identified ($v_2$-related and CME-like ) sources.

Last but not least, let us emphasize also certain ambiguities on the data aspects. In particular there appears to be noticeable difference between the two elliptic flow $v_2$ data sets for AuAu: one  reported in ~\cite{Adams:2004bi} (which was used by us in the decomposition analysis via Eqs.(\ref{dgamma})(\ref{ddelta})), and the other shown in ~\cite{Wang:2012qs} which compared AuAu and UU $v_2$. Such differences may bear certain impact on our extrapolation analysis.
A finalized UU data set with more detailed systematics information on multiplicity, flow, $\g_{\a \b}$ as well as $\d_{\a \b}$ from the experimental side would be highly desirable for further analysis to clarify the situation. One may also notice that there is a ``turning'' behavior on the upper right end of the curve: this corresponds to the region of small multiplicity or peripheral collisions where the $v_2$ has non-monotonic dependence (arising from a competition between increasing initial eccentricity and decreasing density thus smaller pressure gradient that pushes flow) on multiplicity (see e.g. data plot in ~\cite{Wang:2012qs}).

\section {Summary and Discussions}\label{discu}

In summary, we have studied the charge-dependent azimuthal correlations in relativistic heavy ion collisions, as motivated by the search for the Chiral Magnetic Effect (CME) and the investigation of related background contributions. In particular we have attempted to understand how these correlations induced by various proposed effects evolve from collisions with AuAu system to that with UU system.

To do that, we have first systematically studied the electromagnetic fields generated in the RHIC UU collisions
at $\sqrt{s}=193\, \rm GeV$, incorporating initial state fluctuations  on events-by-event basis. The  fluctuations of the
proton position in nucleus cause the fluctuations in the electromagnetic fields, which
on one hand cause sizable electromagnetic field even for some very central events while on the other hand  suppress the correlation between the $\bB$-field orientation and the matter geometry (characterized by event-wise participant plane).
We have quantified the ``projected field strength''  $\lan(e\bB)^2\cos[2(\psi_\bB-\psi_2)]\ran$ (in \fig{fig_b2cos}) which control the $\bB$-field induced effects such as the CME.

We have then used these results to study the recent charge-dependent azimuthal correlation measurements by STAR collaboration for UU collisions at $\sqrt{s}=193 \, \rm GeV$.  Taking the experimental data for charge-dependent azimuthal correlations from AuAu collisions, we have developed a two-component decomposition (see \eq{gab} and \eq{dab}) based on  two types of identified sources ($v_2$-related and possible CME like) that could contribute to the measured correlations.
We have further extrapolated each component to UU system with reasonable assumptions and compared the resulting correlations with data from recent STAR measurements. Based on such analysis we discuss the viability for explaining the data with a  combination of the CME-induced and flow-induced contributions. The extrapolation results have demonstrated similar trend and order-of-magnitude agreement with data while still bearing quantitative discrepancy in the low multiplicity region. We have investigated the uncertainties in the two-component decomposition by varying the factor $\k$ as in the STAR analysis. From these studies it may be concluded that the suggested two-component scenario with $v_2$-related and possible CME-like contributions may be a viable explanation of the measured charge-dependent azimuthal correlations, but also calls for more careful modelings and more detailed experimental information.

We end with a brief discussion on the evolution of these correlations $\g$ and $\d$ with collisional beam energies in light of the two-component scenario. In general one would expect variations of the functions $f$ and $h$ with beam energy, and that is why an extrapolation analysis to other beam energies becomes difficult. For example, going to low energy collisions (as done in RHIC Beam Energy Scan), the physics could change drastically with the ``turning-off'' of a dominant piece of QGP in the fireball evolution. At low enough energy one may expect the disappearance of CME (when the medium is no longer chirally restored) as well as the disappearance of the LCC (when the charge-carriers are not mobile enough to ensure local charge neutrality) --- in that case the $\D \g$ would approach zero as indeed seen in data~\cite{Wang:2012qs}. Going to the higher energy collisions as at LHC, one might expect a qualitatively similar pattern of the multiplicity-dependence of $f$ and $h$ as shown in Fig.\ref{fh}, though possibly with overall magnitude shifting mildly. For the CME-like component, its energy dependence critically depends on the magnetic field: both its magnitude and its duration in time. While the magnitude scales approximately as $\sqrt{s}/M$ (with $M$ the proton mass) and time duration scales in the inverse way as $1/(\sqrt{s}/M)$ so the energy dependence of $\bB$ field may turn out to be rather mild. Going from RHIC to LHC energies, the integrated flow $v_2$ does not change much but the multiplicity at LHC (for same centrality class) does increase a lot (by about a factor $\sim 2$). Noting that the function $h$ (for CME-like source) tends to be saturated for increasing multiplicity while the function $f$ (for flow-related source) tends to be suppressed toward increasing multiplicity, one would then expect that the $\D\g$ at LHC would be quite close to that at RHIC while the $\D \d$ would be somewhat smaller at LHC than at RHIC: both seem to be in line with the ALICE data~\cite{Abelev:2012pa}.

{\bf Acknowledgments:} We are grateful to G. Wang for providing data for UU collisions and for helpful communications. We also thank W. T. Deng and A. Tang for useful discussions. JL is grateful to V. Koch for some early discussions on the decomposition analysis. The research of JL is supported by the National Science Foundation under Grant No. PHY-1352368. JL also thanks the RIKEN BNL Rsearch Center for partial support. X.Z. is currently supported by the US Department of Energy under grant DE-FG02-93ER-40756. X.G.H. is currently under the support of Fudan University grants EZH1512519 and EZH1512600 and Shanghai Natural Science Foundation grant 14ZR1403000.

\end{document}